\documentclass[10pt]{article}
\usepackage{amsmath}
\usepackage{amssymb}
\usepackage{graphicx}
\usepackage{subfigure}
\usepackage{color}
 \usepackage{lineno}
\usepackage{natbib}
\usepackage[margin=1in]{geometry}

 \usepackage[T1]{fontenc} 
\setkeys{Gin}{draft=false}

\newcommand{\beqn}{\begin{equation}}
\newcommand{\eeqn}{\end{equation}}
\newcommand{\beqa}{\begin{eqnarray}}
\newcommand{\eeqa}{\end{eqnarray}}
\newcommand{\beqanonum}{\begin{eqnarray*}}
\newcommand{\eeqanonum}{\end{eqnarray*}}
\newcommand{\beqnonum}{\begin{equation*}}
\newcommand{\eeqnonum}{\end{equation*}}

\newcommand{\bbf}{\begin{bf}}
\newcommand{\ebf}{\end{bf}}
\newcommand{\eqnref}[1]{(\ref{#1})}

\newcommand{\n}{\nonumber}

\newcommand{\der}[2]{\ensuremath{\frac{d #1}{d #2}}}

\newcommand{\ppz}{\ensuremath{\partial_z}}
\newcommand{\ppt}{\ensuremath{\partial_T}}

\newcommand{\cotwo}{\ensuremath{\mathrm{CO_2}}}
\newcommand{\othree}{\ensuremath{\mathrm{O_3}}}

\newcommand{\Qnet}{\ensuremath{Q_\mathrm{net}}}
\newcommand{\FLW}{\ensuremath{F^\mathrm{LW}}}
\newcommand{\FSW}{\ensuremath{F^\mathrm{SW}}}

\newcommand{\Fnet}{\ensuremath{F^\mathrm{net}}}

\newcommand{\cminverse}{\ensuremath{\mathrm{cm^{-1}}}}

\newcommand{\tauk}{\ensuremath{\tau_k}}
\newcommand{\Wmsq}{\ensuremath{\mathrm{W/m^2}}}

\newcommand{\rhov}{\ensuremath{\rho_\mathrm{v}}}

\newcommand{\Rv}{\ensuremath{R_\mathrm{v}}}

\newcommand{\pvstar}{\ensuremath{p^*_{\mathrm{v}}}}

\newcommand{\Ts}{\ensuremath{T_\mathrm{s}}}

\newcommand{\RH}{\ensuremath{\mathrm{RH}}}

\newcommand{\Tlcl}{\ensuremath{T_\mathrm{LCL}}}
\newcommand{\Ttp}{\ensuremath{T_\mathrm{tp}}}

\newcommand{\Kinverse}{\ensuremath{\mathrm{K^{-1}}}}
\newcommand{\Kelvin}{\ensuremath{\mathrm{K}}}

\newcommand{\kmax}{\ensuremath{k_\mathrm{max}}}
\newcommand{\kmin}{\ensuremath{k_\mathrm{min}}}

\begin{document}

%
%

\title{Invariant Radiative Cooling and Mean Precipitation Change}

%
%

 \author{Nadir Jeevanjee\footnote{Department of Geosciences, Princeton University, Princeton NJ 08544 USA. nadirj@princeton.edu (corresponding author)} \footnote{Princeton Program in Atmosphere and Ocean Sciences, Princeton University, Princeton NJ 08540 USA} \footnote{Geophysical Fluid Dynamics Laboratory,  Princeton NJ  08540 USA}\ \    and David Romps\footnote{Department of Earth and Planetary Sciences, University of California at Berkeley, Berkeley, CA 94702  USA.} \footnote{Climate and Ecosystems Science Division, Lawrence Berkeley National Laboratory, Berkeley, CA USA}
}

\maketitle

\begin{abstract}
We show that radiative cooling profiles, when described in temperature coordinates, are insensitive to surface temperature \Ts. We argue this theoretically as well as confirm it in cloud-resolving simulations of radiative convective equilibrium (RCE). This \Ts-invariance holds for shortwave and longwave cooling separately, as well as their sum. Furthermore, the \Ts-invariance of radiative cooling profiles leads to a simple expression for the \Ts-dependence of column-integrated cooling and hence precipitation, and gives insight into why mean precipitation increases at a rate of $2 -3\%\ \Kinverse$ in RCE.  The relevance of these results to global climate simulations is assessed, and the \Ts-invariance is found to hold in the mid and upper troposphere.  In the lower troposphere, the pressure-invariance of cloud layers and circulation tends to dominate.
%
%
\end{abstract}

%
%

\section {Introduction}
Despite its fundamental role in driving atmospheric motions, atmospheric radiative cooling remains somewhat enigmatic. Though the fundamentals of radiative transfer are quite well-understood and have been for some time, translating these fundamentals into realistic cooling rates requires a symphony of complicated spectroscopic and radiative transfer calculations which render the final result somewhat inscrutable. As a result, we lack simple descriptions of the radiative cooling profiles produced by our numerical models.

One implication of this is that quantities that are closely tied to radiative cooling, such as global mean precipitation, also remain somewhat enigmatic. We do know that the atmospheric (rather than planetary) energy budget, in which condensation heating from precipitation balances atmospheric radiative cooling, constrains global mean precipitation $P$ to be roughly equal to column-integrated net radiative cooling $\Qnet$ \citep{ogorman2012,allen2002}:
\beqn
	LP = \Qnet \quad \mbox{(\Wmsq)} \label{rad_precip_constraint}
\eeqn
(here $L$ is the latent heat of vaporization). We also know that both cloud-resolving models and global climate models robustly exhibit  mean precipitation increases with warming of $2 -3\%\ \Kinverse$  \citep{stephens2008a, lambert2008, held2006}. Furthermore, \cite{pendergrass2014} recently explained this increase in global mean precipitation in terms of an increase in downward radiative emission from the atmosphere at the surface. Despite this progress, however, a basic question remains unanswered: why does this increase take on the value that it does? Why $2 -3\%\ \Kinverse$ and not many times  larger or smaller?

This paper aims to reveal some simple behavior in radiative cooling profiles, and to use it to  answer this question about precipitation change. The physics we rely on is not new, but was rather noted as far back as \cite{simpson1928}, and revived recently by \cite{ingram2010}. Our contribution here is to shift the focus from outgoing longwave radiation to atmospheric radiative cooling, and to extend the argument to both the longwave (LW) and shortwave (SW) channels. 

We will  focus on how vertically-resolved radiative cooling profiles change with warming, rather than focusing on radiative fluxes at the surface or TOA as in \cite{pendergrass2014}. In particular, we will argue, following \cite{simpson1928} and \cite{ingram2010}, that water vapor density and optical depth profiles should behave very simply when considered as functions of temperature as a vertical coordinate. This then implies almost immediately that LW and SW radiative flux divergences should also behave simply in temperature coordinates. This simple behavior leads to a prognostic expression for $d\Qnet/d\Ts$ and hence $dP/d\Ts$ (\Ts\ is surface temperature), which we validate with a cloud-resolving model (CRM). We then seek insight from this framework, and then ask to what extent our CRM results generalize  to global climate models (GCMs).

\section{CRM Simulations of RCE}
We begin by studying precipitation change in one of the simplest systems in which the radiative constraint on precipitation \eqnref{rad_precip_constraint} operates, namely tropical oceanic radiative-convective equilibrium (RCE) with fixed sea-surface temperature. This system approximates the real tropics, where the majority of Earth's precipitation occurs \citep{simpson1988}, and like the GCMs exhibits precipitation increases of roughly $2 -3\%\ \Kinverse$ \citep{romps2011, muller2011b}.  

We simulate RCE using Das Atmosph\"arische Modell \citep[DAM,][]{romps2008},   a three-dimensional, fully-compressible, non-hydrostatic cloud-resolving model, coupled to radiation via the Rapid Radiative Transfer Model 
\citep[RRTM,][]{mlawer1997}. DAM employs the six-class Lin-Lord-Krueger  microphysics scheme \citep{lin1983, lord1984, krueger1995}, and relies on implicit LES \citep{margolin2006} for sub-grid scale transport, so no explicit sub-grid scale turbulence scheme is used.

\begin{figure}[t]
	\begin{center}
			\includegraphics[scale=0.5]{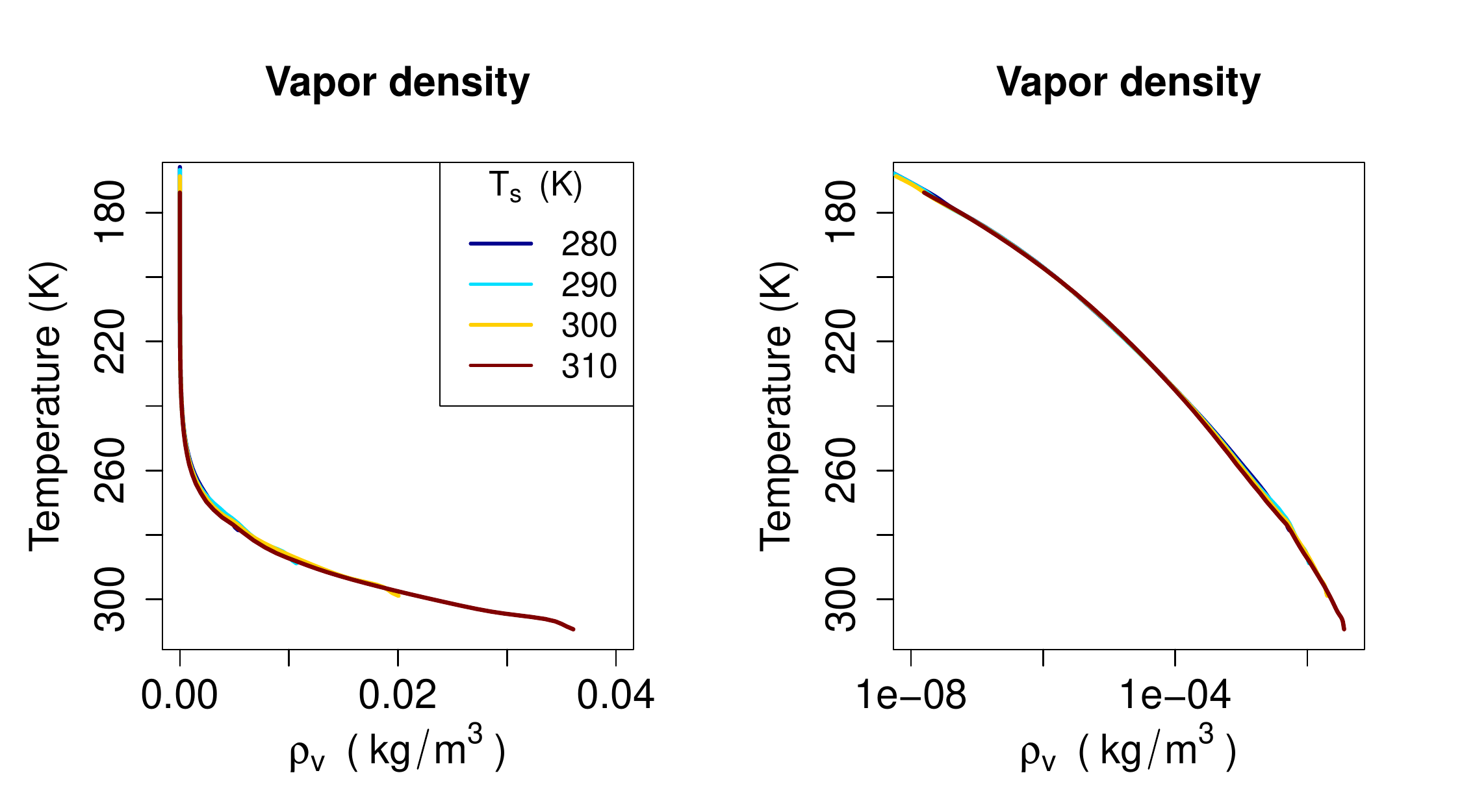}
		\caption{Profiles of $\rhov(T)$ from our RCE simulations at various \Ts, with both linear and log scales. These profiles are `\Ts-invariant' in the sense that $\rhov(T)$ does not depend on \Ts, i.e. that the \rhov\ profiles at different \Ts\ collapse onto a single curve.
		\label{rhov_fig}
		}
	\end{center}
\end{figure}

	Our RCE simulations ran on a square doubly-periodic domain of horizontal dimension $L=72$ km, with  a horizontal resolution of $dx=1$ km. The vertical grid stretched smoothly from 50 m resolution below 1000 m to 250 m resolution between 1000 m and 5000 m, and then to 500 m up to the model top at  30 km. We calculated surface heat and moisture fluxes using a bulk aerodynamic formula, and used a pre-industrial \cotwo\  concentration of 280 ppm with no ozone except where specified otherwise. To explore precipitation changes  with warming we ran five experiments at surface temperatures of $\Ts=(280,290,300,310,320)$ K. Our runs branched off the equilibrated runs described in \cite{romps2014}, and were run for 60 days  to iron out any artifacts from changing the domain and resolution. All vertical profiles are time-mean and domain-mean, averaged over the last 20 days of each run. 

Since we run with prescribed \Ts, our warming experiments are somewhat artificial, in that the warming is not driven by increases in \cotwo. This has the advantage that we isolate part of the physics and thus have a better chance at arriving at a simple description, but has the disadvantage that we omit the direct effect of increased \cotwo\ on atmospheric cooling and hence precipitation, an effect of roughly -1 \Wmsq/K \citep{pendergrass2014}. This omission does not affect our main conclusions about precipitation change.

\section{\Ts-invariance of Flux Divergences}
\label{Ts_invariance}
The simple behavior of radiative cooling alluded to above begins with the key fact that  the water vapor density 
	\beqn
		\rhov =  \RH\frac{\pvstar(T)}{\Rv T} \; 
	\label{rhov}
	\eeqn
	 is (up to variations in relative humidity \RH) a function of temperature only. [Note that it has been shown recently that RH is itself a function of $T$ in RCE \citep{romps2014}. Also note that here $p_v^*$  is the saturation vapor pressure of water, and all other symbols have their usual meaning.] If we use $T$ as a vertical coordinate,  Eqn. \eqnref{rhov} then tells us that the function $\rhov(T)$ does not depend on \Ts. This is what we mean by `\Ts-invariance'. We verify \Ts-invariance of $\rhov(T)$  in Fig. \ref{rhov_fig}, where indeed  the \rhov\ profiles at different \Ts\ collapse onto a single curve when plotted in temperature coordinates.

\begin{figure}[t]
	\begin{center}
			\includegraphics[scale=0.5]{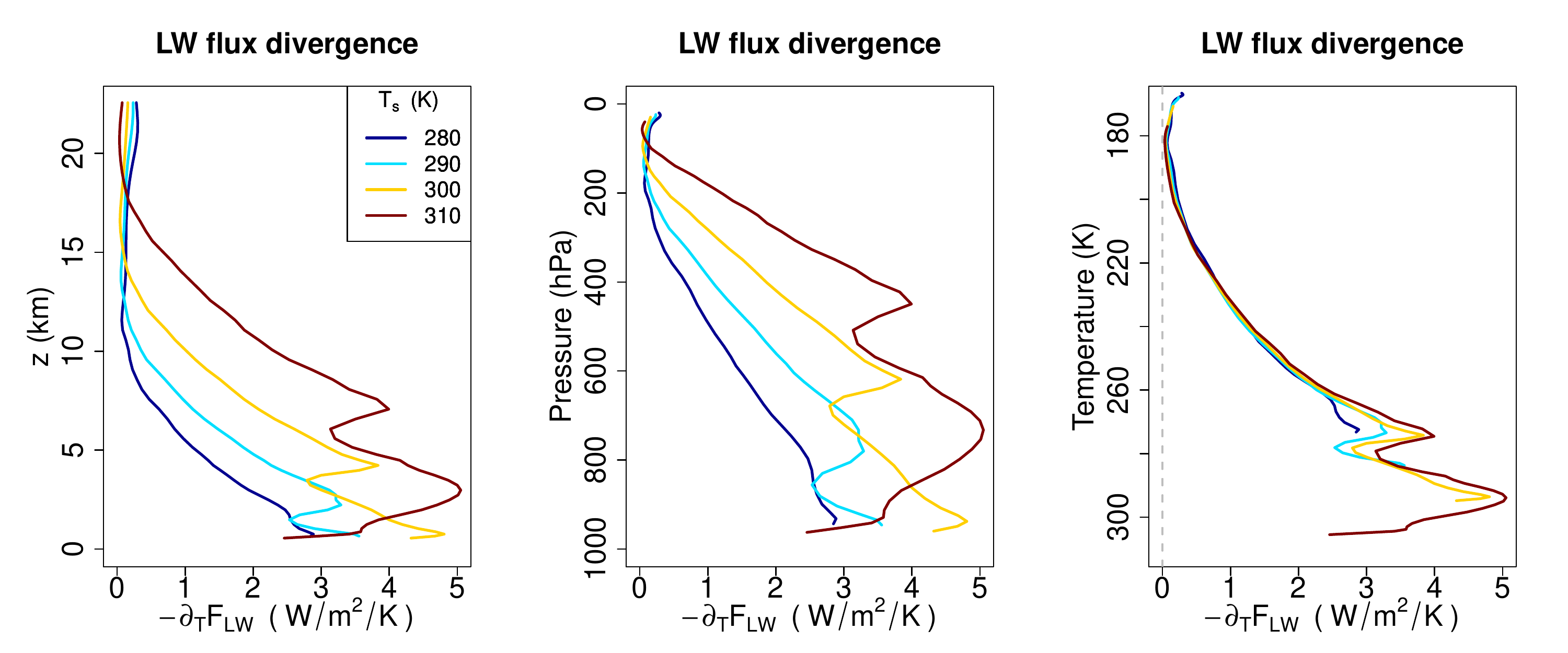}
		\caption{LW flux divergence  $-\ppt \FLW$, as diagnosed from RRTM coupled to our CRM RCE simulations at \Ts=(280,\ 290,\ 300,\ 310) K (the 320 K simulation is omitted for clarity). Fluxes are plotted from the lifting condensation level of each simulation to 22.5 km for clarity, and  in height, pressure, and temperature coordinates to emphasize the \Ts-invariance of  $(-\ppt \FLW)(T)$. The gray dotted line in the right panel plots $-\ppt \FLW = 0$, and shows the \Ts-invariance of $\Ttp \approx 185$ K.
		\label{pptflw_tinv_dam}
		}
	\end{center}
\end{figure}

\begin{figure}[t]
	\begin{center}
			\includegraphics[scale=0.5]{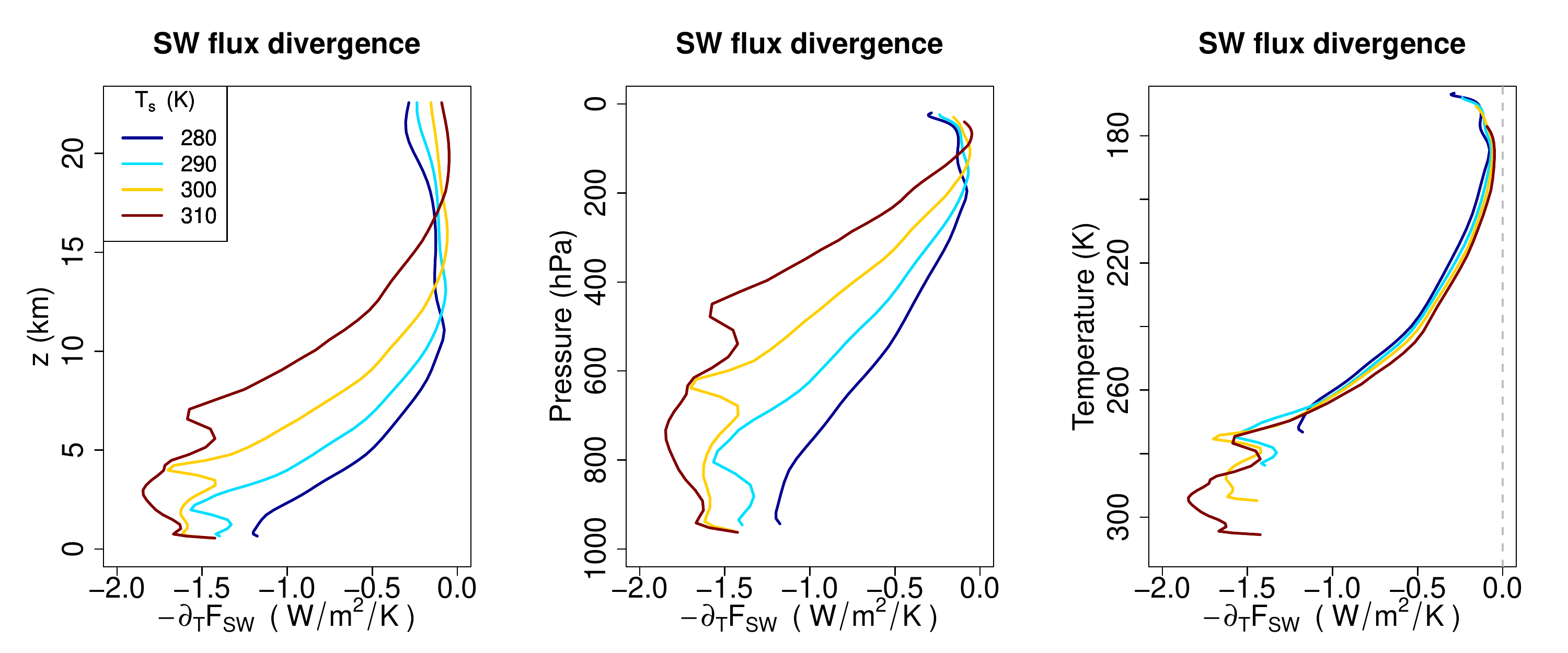}
		\caption{As in Fig. \ref{pptflw_tinv_dam}, but for SW instead of LW.
		\label{pptfsw_tinv_dam}
		}
	\end{center}
\end{figure}

	For wavenumbers $k$ outside of spectral bands where other trace gases (like \cotwo\ and \othree) dominate, the optical depth $\tauk$ is just
	\beqn
		\tau_k(z) = \int_z^\infty \kappa(k)  \rhov(z') \, dz'  \; 
		\label{tauz}
	\eeqn
		where $\kappa(k)$ is a  mass absorption coefficient  (units $\mathrm{m^2/kg}$) whose pressure-broadening and temperature scaling we neglect. Changing the integration variable to temperature $T'$ yields
		\beqn
		\tau_k(T) \approx \int_{\Ttp}^T \kappa(k)  \rhov(T') \, \frac{dT'}{\Gamma}  \; ,
		\label{tauT}
	\eeqn
	where we neglect stratospheric water vapor and take the lower limit of the integral to be the tropopause temperature $\Ttp \approx 185$ K, where radiative cooling goes to 0 (see Figs. \ref{pptflw_tinv_dam} and  \ref{pptfsw_tinv_dam}, which also show that \Ttp\ is \Ts-invariant). The only quantity in Eqn. \eqnref{tauT} that might still exhibit some \Ts-dependence is the  moist lapse rate $\Gamma$, but Figure 2 of \cite{ingram2010} shows that when $\Gamma$ is considered a function of temperature, it too is fairly  \Ts-invariant. Equation \eqnref{tauT} then implies that \tauk\ profiles at any $k$ exhibit the same \Ts-invariance as \rhov. This argument was also made by \cite{ingram2010}, and its essence goes back to  \cite{simpson1928}.
	
	To connect all this with radiative cooling, we invoke the cooling-to-space  approximation \citep[e.g.,][]{thomas2002, rodgers1966}, which says that the spectrally resolved LW flux divergence in temperature coordinates $-\ppt \FLW_k$ (units $\Wmsq/\mathrm{K}/\cminverse$, minus sign introduced to maintain a consistent sign with  $\ppz \FLW_k$) is approximately
	\beqn
		-\ppt \FLW_k \approx - \pi B_k(T) \frac{d (e^{-\tauk(T)})}{dT} \ ,
	\label{cts_spectral}
	\eeqn
where  the transmission function $e^{-\tauk}$ gives the fraction of radiation emitted at a given height that travels unabsorbed out to space. Since the Planck function $B_k(T)$ is \Ts-invariant, and $\tauk(T)$ is as well, we also expect $-\ppt \FLW_k$ to be \Ts-invariant. Since this holds for all $k$ where water vapor dominates, it should also hold approximately for the spectrally integrated LW flux divergence $-\ppt \FLW$ ($\Wmsq/\mathrm{K}$). This is confirmed in  Fig.  \ref{pptflw_tinv_dam}, which plots $(-\ppt \FLW)(T)$ as diagnosed from RRTM coupled to our  RCE simulations.  That figure also plots $-\ppt \FLW$ as functions of $z$ and $p$, to emphasize that this invariance only holds  when $T$ is used as the vertical coordinate.

	A similar argument holds for the SW flux divergence. If $I_k$ is the incident solar flux at wavenumber $k$, and  neglecting reflection and scattering in the  near-infrared, 
then without further approximation we have
	\beqn
		-\ppt \FSW_k = - I_k \der{(e^{-\tauk(T)})}{T}
		\
	\eeqn
\citep[c.f.][eqn. 9.26]{thomas2002}. This equation is similar to  \eqnref{cts_spectral} but with $B_k(T) \rightarrow I_k$, and since $I_k$ is also \Ts-invariant, we can argue as above that $(-\ppt \FSW)(T)$ should be \Ts-invariant. This is confirmed in Fig. \ref{pptfsw_tinv_dam}, where again the simple behavior of $-\ppt \FSW$ in temperature coordinates is contrasted with that in height and pressure coordinates.

The fluxes used in Figs.  \ref{pptflw_tinv_dam} and \ref{pptfsw_tinv_dam} are all-sky fluxes, but the foregoing argument was for clear-sky fluxes. This is permissible because cloud fractions in our RCE simulations are low (attaining a maximum of $\sim 10 \%$ at the anvil height in our simulations), so it is the clear-sky physics which dominates. We will touch upon cloud radiative effects in section \ref{sec_GCMs}, when we assess how well these CRM results generalize to GCMs.
 
		
\section{A simple picture for column-integrated radiative cooling} \label{sec_simple_Q}

Now that we have established  the \Ts-invariance of radiative flux divergences, we can construct a simple, quantitative picture of how column-integrated radiative cooling, and hence precipitation,  changes with surface temperature. 
	
	Let $F$ denote radiative flux in a particular channel -- LW, SW, or Net (LW+SW) -- and $Q$ the associated column-integrated free-tropospheric radiative cooling. We consider  the free troposphere, rather than the full troposphere, because the radiative constraint on precipitation 
		\beqn
			LP \approx \Qnet
		\label{p_constraint}
		\eeqn
		 holds best for the free troposphere  \citep{ogorman2012}. We define the free troposphere here as being above the lifting condensation level \Tlcl\ and below the tropopause \Ttp.
	
\begin{figure}[t]
	\begin{center}
			\includegraphics[scale=0.5,trim=0cm 0cm 0cm 5cm,clip=true]{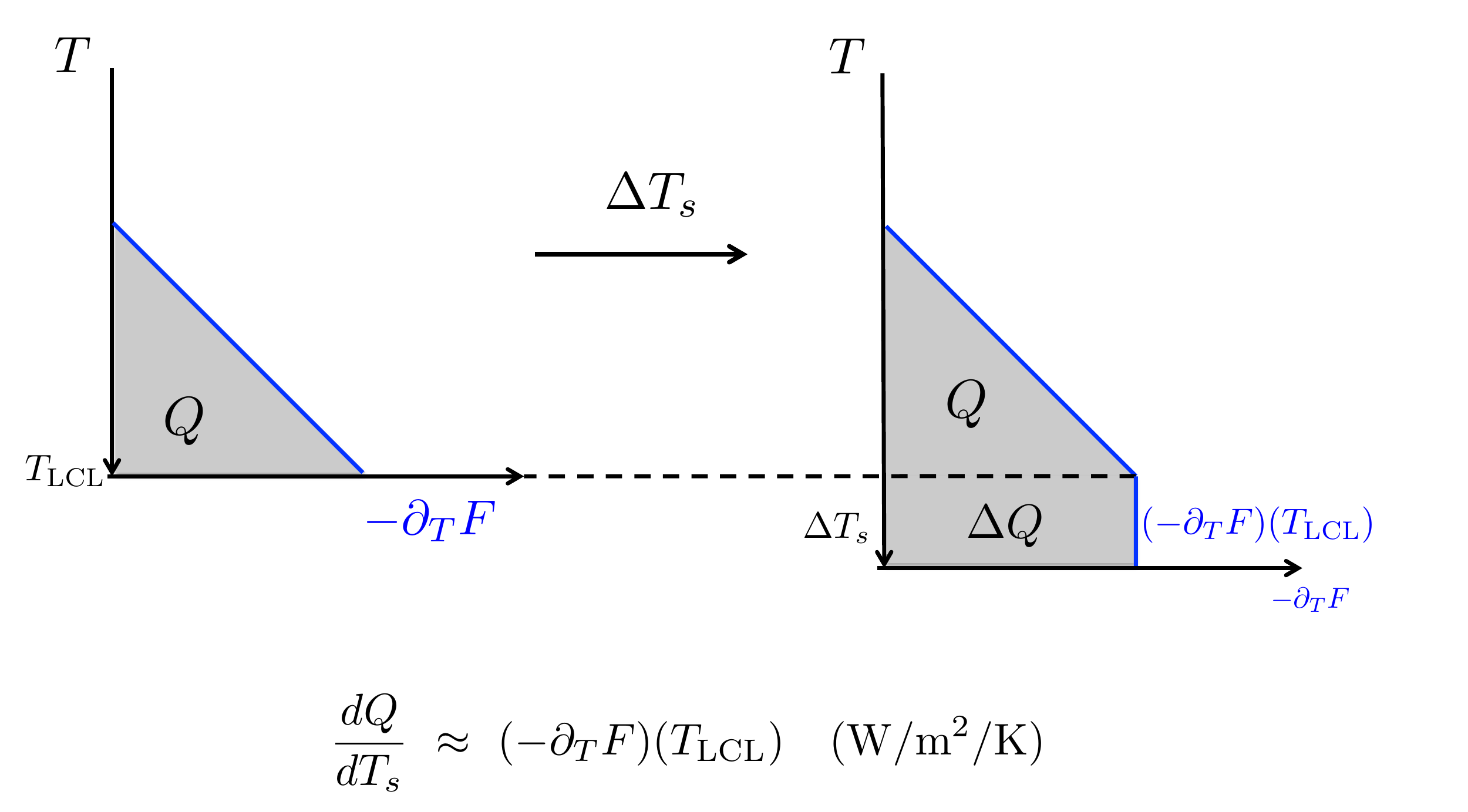}
		\caption{Cartoon depicting the increase in $Q$ with \Ts\ in Eqn. \eqnref{dqdts}. Increasing the temperature range of the troposphere  exposes more of the \Ts-invariant curve $(\ppt F)(T)$ (blue lines). The contribution  of this newly exposed region to column-integrated cooling is given by Eqn. \eqnref{dqdts}.
		\label{dqdts_cartoon}
		}
	\end{center}
\end{figure}

	  The basic idea is to write $Q$ as an integral of $-\ppt F$  in temperature coordinates: 
	\beqn
		Q =  \int_{\Ttp}^{\Tlcl} (-\partial_{T'} F) dT' \ . 
		\n
	\eeqn
   If we approximate the change in  \Tlcl\ as equal to the change in \Ts, then the change in $Q$ with surface temperature is  simply
	\beqn
		\der{Q}{\Ts} \ =\  \left.  -\ppt F\right|_{\Tlcl}  \; .
	\label{dqdts}
	\eeqn
In other words, since the tropospheric cooling profile $(-\ppt F)(T)$  is independent of \Ts, increasing \Ts\ just exposes more of this profile.  The contribution of this new section of the $(-\ppt F)(T)$ curve to $Q$ is given by \eqnref{dqdts}.  A cartoon of this argument is given in Fig. \ref{dqdts_cartoon}. For finite changes in \Ts, Eqn. \eqnref{dqdts} approximates $(-\ppt F)(T)$ in the newly exposed region as equal to $-\ppt F$ at the LCL of the base state, but for small enough changes in \Ts\ this approximation should be adequate. Specializing Eqn. \eqnref{dqdts} to the Net channel and invoking \eqnref{p_constraint} then yields a \emph{prognostic} equation for precipitation change with surface warming.

\begin{figure}[t]
	\begin{center}
			\includegraphics[scale=0.5]{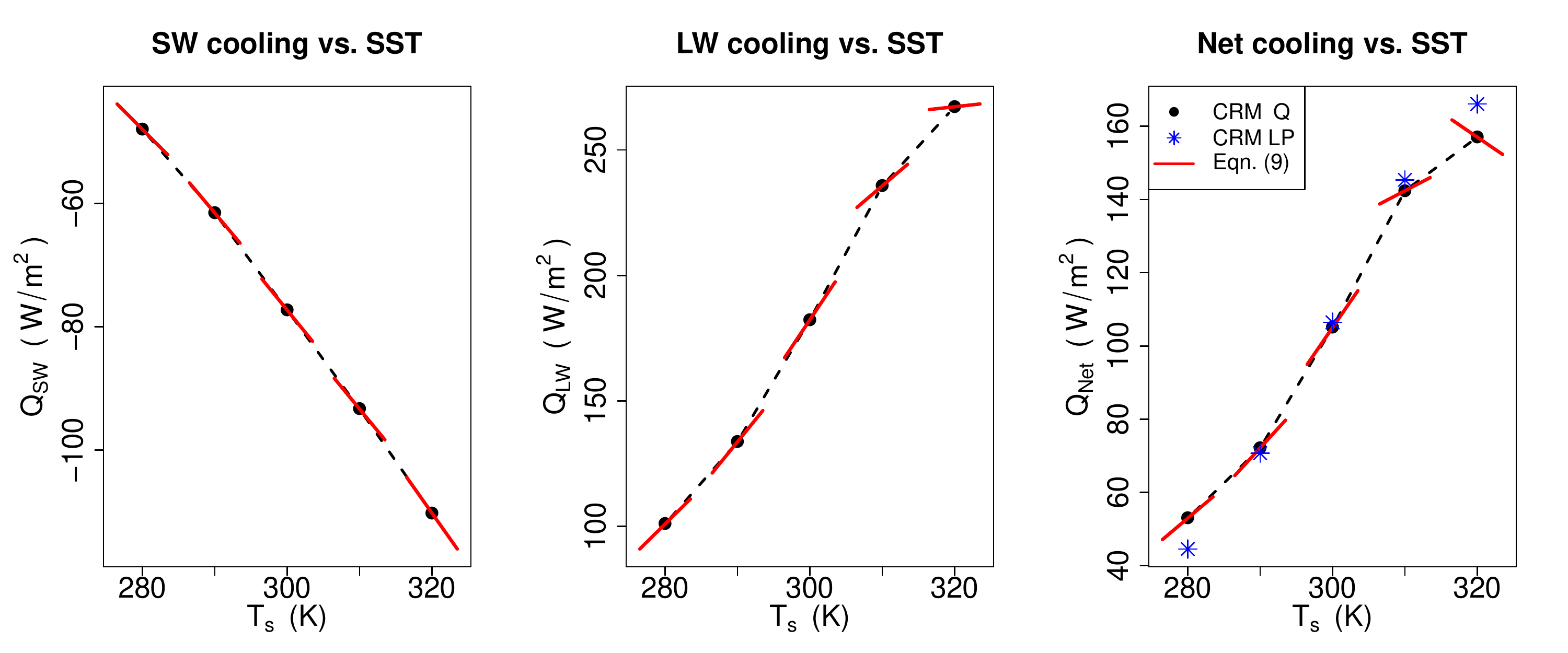}
		\caption{Column-integrated cooling $Q$ vs. \Ts\ (black circles), along with slopes $d Q/d \Ts$ (red lines) as diagnosed from \eqnref{dqdts}. These are shown for the SW (left), LW (center) and Net (right) channels.  The black dashed lines connect the black circles and give a benchmark slope against which to compare the red lines. The `Net' panel also gives CRM-diagnosed precipitation values in blue stars. See text for discussion.
		\label{Qnet_varsst}
		}
	\end{center}
\end{figure}

Let us test the predictive power of Eqn. \eqnref{dqdts}. The panels of Fig. \ref{Qnet_varsst} plot $Q(\Ts)$ as diagnosed directly from our CRM simulations, along with estimates of the slope of this curve diagnosed via  Eqn. \eqnref{dqdts}, for the SW, LW, and Net  channels (\Tlcl\ is diagnosed as $T$ at the low-level maximum in cloud fraction). Precipitation $P$ is also plotted alongside $\Qnet$.  Figure \ref{Qnet_varsst} shows that  Eqn. \eqnref{dqdts}  captures the changes in  cooling in all channels. Furthermore, since $P$ tracks \Qnet\ closely for $290\leq \Ts \leq 310$ K, Eqn. \eqnref{dqdts} also captures precipitation changes, at least in this temperature regime.

We also see that  Eqn. \eqnref{dqdts} predicts a \emph{decrease} in  \Qnet\ with \Ts\ at \Ts=320 K; this is not an error in our diagnostic equation \eqnref{dqdts} , but rather a real effect due to the fact that $-\ppt \FLW$ tends towards zero with increasing $T$  while -\ppt \FSW\ is staying roughly constant. (This behavior of $-\ppt \FLW$ is likely related to runaway greenhouse physics, known to set in at roughly 310 K \citep{goldblatt2013}.) This leads to radiative heating, rather than cooling, in the  lower troposphere, which violates the basic radiative-convective paradigm; it is perhaps then no surprise that the constraint \eqnref{p_constraint} appears to break down in this \Ts\ regime. An analogous high \Ts\ breakdown of the radiative constraint on precipitation can also be found in energetically consistent experiments \citep{lehir2009, pierrehumbert1999}.  The radiative  constraint  also breaks down at low \Ts\ (i.e. $\Ts \leq 280$ K), where sensible heat fluxes start to dominate over latent heat fluxes. Thus, Eqn. \eqnref{dqdts} has explanatory power for  precipitation changes at  temperatures somewhat greater than or equal to Earth's mean temperature of 288 K. Outside the $290\leq \Ts \leq 310$ K range, other constraints besides our purely radiative one seem to be required to predict changes in $P$.

\section{Why does precipitation increase at $2 -3\%\ \Kinverse$?} \label{sec_1percent}
The results in Fig. \ref{Qnet_varsst} show that our framework  has some predictive power for explaining changes in \Qnet\ and hence $P$ in RCE. Let us then try to use this framework to answer the question posed in the introduction, namely: why does mean precipitation increase at $2 -3\%\ \Kinverse$?

First, let us confirm in a back-of-the-envelope fashion that Eqn. \eqnref{dqdts} indeed gives a $2 -3\%\ \Kinverse$ increase in $P$. Combining \eqnref{p_constraint} and \eqnref{dqdts} gives
	\beqn
		\frac{d \ln  P}{d \Ts} \ \approx\  \frac{(-\ppt \Fnet)(\Tlcl)}{\Qnet} \; .
	\label{precip_estimate}
	\eeqn
For \Ts=300 K, where $(-\ppt \Fnet)(\Tlcl) \approx 3 \ \Wmsq/\mathrm{K}$ and $\Qnet =  104\ \Wmsq$, we find $\frac{d \ln  P}{d \Ts}=  3\%\ \Kinverse$, as expected.

Now, suppose we take \Ts=300 K and  try to simply parametrize the net cooling as $-\ppt \Fnet \propto (T-\Ttp)^\beta$.  Further suppose (motivated by inspection of Figs. \ref{pptflw_tinv_dam} and \ref{pptfsw_tinv_dam})  that $\beta \approx 2$, i.e. that $-\ppt \Fnet$ is roughly quadratic  in $(T-\Ttp)$. Then the full tropospheric radiative cooling is $Q\sim (\Ts-\Ttp)^{\beta+1}$, and hence 
	\beqn
		\frac{d \ln Q}{d \Ts}  =  \frac{\beta+1}{\Ts-\Ttp}\ . \label{dqdts_approx}
	\eeqn
Note that $\Ts-\Ttp$ is the \emph{depth of the troposphere expressed in temperature coordinates}. For  \Ts= 300 K this depth is roughly 100 K, and so \eqnref{dqdts_approx} gives roughly 3 \% \Kinverse, consistent with the result from Eqn. \eqnref{precip_estimate}.

On the other hand, if $-\ppt \Fnet$ were constant throughout the depth of the troposphere, i.e. $\beta=0$, then $Q$ would just scale with $\Ts-\Ttp$. But then it is clear that, since \emph{a 1 K increase in \Ts\  is a $1\%$ increase in tropospheric depth \Ts-\Ttp}, $Q$ should increase at 1 \% \Kinverse. The fact that Q increases somewhat faster than that can then be understood as a result of the fact that $-\ppt \Fnet$ is increasing, not constant, with $T$, i.e. that $\beta>0$ in Eqn. \eqnref{dqdts_approx}.


\section{Extension to GCMs} \label{sec_GCMs}
The framework presented so far has an appealing simplicity. But, the question remains as to whether our results generalize from RCE to much more realistic GCM simulations. Before trying to predict precipitation change in GCMs,  we must first check  whether \Ts-invariance holds in GCMs, in some sense. We do this by binning  GCM columns by their local \Ts, computing an average $-\ppt F$ profile for each bin, and then checking the \Ts-invariance of each of these profiles. 

\begin{figure}[t]
	\begin{center}
			\includegraphics[scale=0.7]{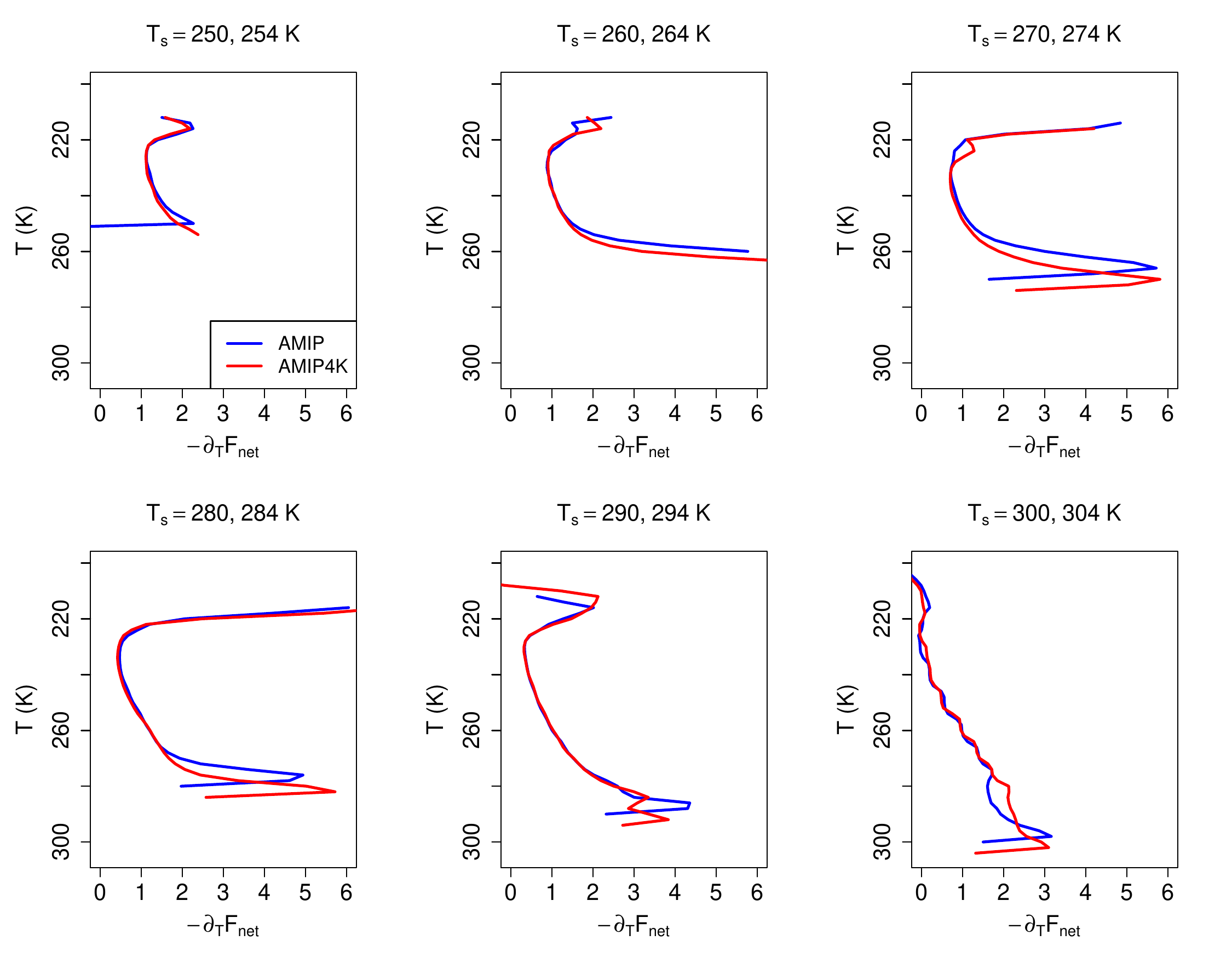}
		\caption{ Profiles of $-\ppt \Fnet$ for various \Ts\ bins for the AMIP and AMIP4K runs of IPSL-CM5A-LR.  These profiles show that \Ts-invariance holds in this GCM in the mid and upper troposphere, but not near the surface (i.e. $T \lesssim \Ts$).
		\label{fnet_ipsl}
		}
	\end{center}
\end{figure}

For this we utilize the AMIP and AMIP4K  output in the CMIP5 archive. These experiments are atmosphere-only, and feature observed SSTs (AMIP) as well as uniform perturbations to those observed SSTs (AMIP4K), with no change in \cotwo\ concentration; as such they are good analogs to our CRM experiments. The AMIP4K experiment was part of the CFMIP protocol, which also requested the output of vertically-resolved radiative fluxes, rather than just surface and TOA fluxes, allowing us to compute $-\ppt F$ profiles.

Six models participated in the AMIP and AMIP4K CFMIP experiments and provided the output we require. We begin by analyzing one of them, IPSL-CM5A-LR. Figure \ref{fnet_ipsl} shows  profiles of average $-\ppt \Fnet$ for six of our \Ts\ bins, where for each \Ts\ bin the average is taken over  all columns from the last 30 years of the simulation for which the lowest model-level air temperature lies in the range $(\Ts,\Ts +2\Kelvin)$. For the AMIP4K calculation in each panel the $\Ts +4\Kelvin$ bin is used, so as to compare roughly the same columns between the two simulations. More details on this calculation are given in the Appendix, which also shows the decomposition of these profiles into their LW and SW components. 

\begin{figure}[t]
	\begin{center}
			\includegraphics[scale=0.7]{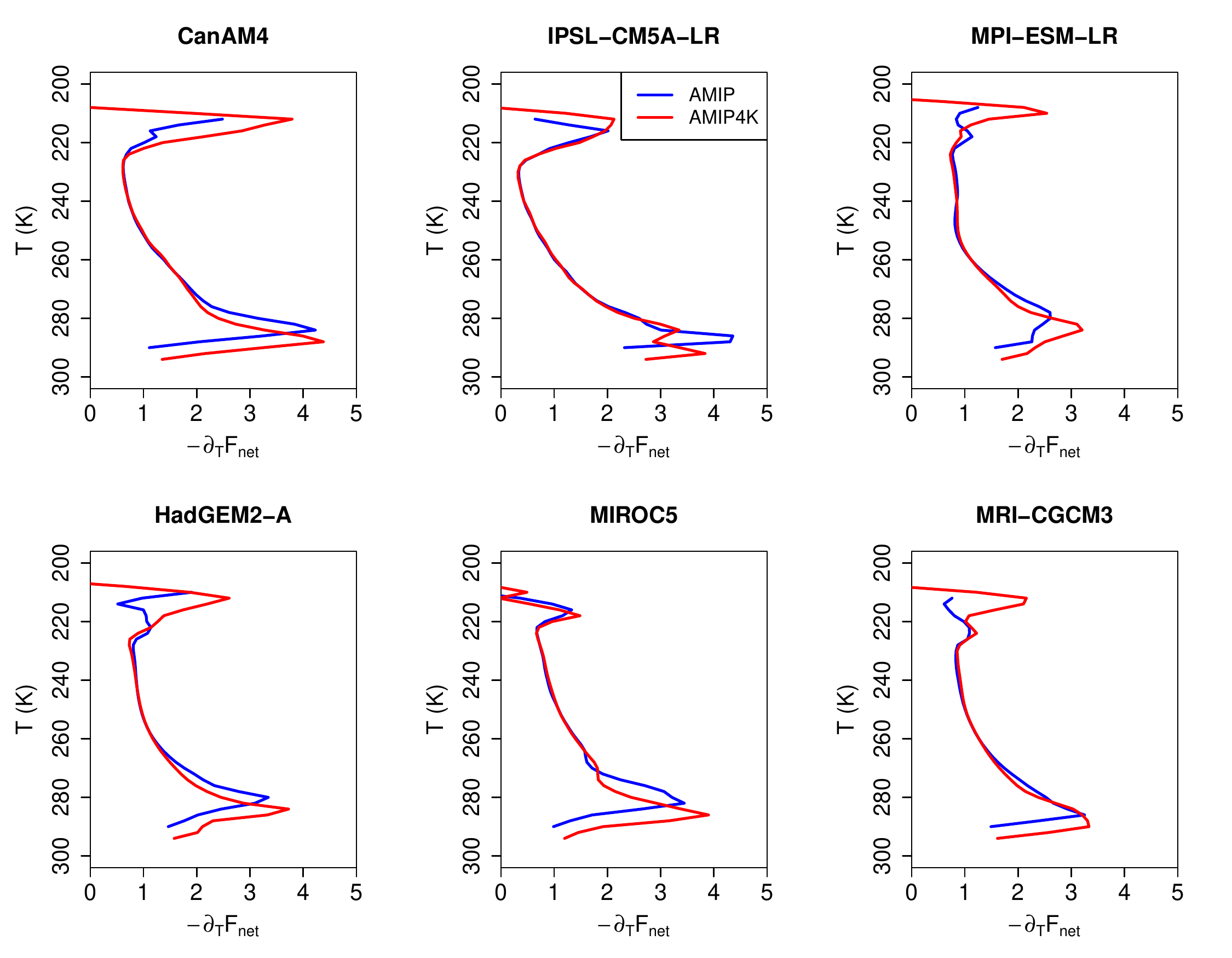}
		\caption{ Profiles of $-\ppt \Fnet$ for the \Ts=290 K (AMIP) and 294 K (AMIP4K) bins for all six CFMIP models. These profiles show that \Ts-invariance in the mid and upper troposphere holds across models.
		\label{fnet_all}
		}
	\end{center}
\end{figure}

The take-away from Figure \ref{fnet_ipsl} is that for a given \Ts, \Ts-invariance seems to hold quite well in the mid and upper troposphere, but not in the lower troposphere. This appears to be due to cloud and circulation  effects, such as inversion layers and their associated low clouds, which appear to stay at a fixed pressure (rather than temperature) with warming. This yields features in the $-\ppt \Fnet$ profiles that appear to shift down to higher temperatures with  warming. This stands in contrast to the CRM $-\ppt F$ profiles, which under warming reach the new \Ts\ not by \emph{shifting} down but by \emph{extending} downward. 

Such a qualitative difference between the GCMs and our CRM means that Eqn. \eqnref{dqdts} cannot be applied to the GCMs, at least not without modification. At the same time, we do see that \Ts-invariance holds throughout much of the atmosphere for many \Ts\ regimes, and may thus still be a useful approximation for other problems. Of course, Fig. \ref{fnet_ipsl} only establishes this for one model, so robustness across models still needs to be checked. We do this in Fig. \ref{fnet_all}, which shows average $-\ppt \Fnet$ profiles for the $\Ts= 290$ (AMIP) and 294 (AMIP4K) bins  for all six of our  CFMIP models. These panels show that the IPSL model is not an outlier, and that \Ts-invariance in the mid and upper troposphere is robust across models.

\section{Summary and Discussion}
We summarize our findings as follows:
	\begin{itemize}
		\item Radiative cooling profiles in temperature coordinates are  \Ts-invariant. This \Ts-invariance holds for the shortwave and longwave separately, as well as together (Figs. \ref{pptflw_tinv_dam}, \ref{pptfsw_tinv_dam}).
		\item For RCE, this \Ts-invariance yields a simple model for how column-integrated cooling and  precipitation change with \Ts\ [Eqn. \eqnref{dqdts}]. This model captures the simulated changes (Fig. \ref{Qnet_varsst}), and also leads to an even simpler model [Eqn. \eqnref{dqdts_approx}] which yields insight into why precipitation changes are  $2 -3\%\ \Kinverse$ in RCE.
		\item For \Ts-binned $-\ppt F$ profiles from GCMs, \Ts-invariance holds in the mid and upper troposphere, but not near the surface (Figs. \ref{fnet_ipsl}, \ref{fnet_all}).
	\end{itemize}
		
An obvious question left unanswered here is whether the $2 -3\%\ \Kinverse$ increase in precipitation found in GCMs is at all physically analogous to that in CRMs, or whether different physics is at play. Inspection of Fig. \ref{fnet_all} shows that the near-surface peak in $-\ppt \Fnet$ appears to stay at a fixed pressure with surface warming, but also increases in magnitude. Presumably this is due to an increase in (either cloudy or clear-sky) Planck emission, but further work would be needed to check this and model it in such a way as to give a prognostic expression for $d\Qnet/d\Ts$. 

There are also unanswered questions regarding the  argument given in Section \ref{Ts_invariance}. For instance, to what degree are optical depth profiles for water vapor lines actually \Ts-invariant, as claimed here and by \cite{ingram2010}? Would a line-by-line calculation verify this?  Also, what are the conditions for the cooling-to-space approximation to be valid? And finally, why does  the radiative tropopause temperature \Ttp\ appear to be fixed in our simulations? This bears a certain resemblance to the FAT hypothesis but is distinct from it, as the radiative tropopause and anvil peak are distinct features of the atmosphere and occur at quite different heights (approximately 17 km and 11 km, respectively, in the present day deep tropics).

There is also the question of robustness of our RCE results to choice of CRM. While CRMs do not employ as many parameterizations as GCMs, they must still choose sub-grid turbulence and microphysics schemes, which can lead to substantial uncertainties in some variables including cloud cover \citep[e.g.][]{tsushima2015, igel2014}. Since the arguments given here were clear-sky arguments and relied on the low values of  cloud fraction exhibited by DAM, it is possible that the \Ts-invariance exhibited here may not hold as well in other CRMs. The upcoming RCE Model Intercomparison Project \citep[RCEMIP,][]{wing2017b} would make an ideal venue for investigating this.

Finally, we should note that the $7\%\ \Kinverse$ Clausius-Clapeyron scaling of $\pvstar(T)$ plays no role in setting the $2 -3\%\ \Kinverse$  scaling of \Qnet\ and $P$. This can be seen most directly by appealing to Eqn. \eqnref{dqdts_approx}, which is a simple consequence of the \Ts-invariance of -\ppt \Fnet\ and \Ttp. The \Ts-invariance of -\ppt \Fnet\  follows from  $\pvstar(T)$ being a function of temperature only, with no requirement that $p_v^*(T)$ even be exponential in $T$, let alone that $d\ln \pvstar/dT \approx 0.07 \ \Kinverse$.  Our arguments thus suggest that this value could be doubled or halved without affecting the scaling of $P$. Thus, the Clausius-Clapeyron and mean precipitation scalings should be thought of as independent constraints, one thermodynamic and one radiative, with different physical origins. That they are independent and may thus be combined without circularity is what makes them powerful, allowing for e.g. a prediction of how convective mass fluxes change with warming \citep[][]{held2006}.


\section{Appendix}
This appendix describes in detail our calculation of bin-averaged flux divergence profiles from GCM output.

For a GCM column at a given longitude, latitude, and time, we must first identify a range of tropospheric model levels $k$ over which the temperature $T$ varies monotonically. We identify the uppermost of these levels \kmax\ as the minimum  $k>10$ for which 
$T[k+1]>T[k]$. If none such exists (i.e. no stratospheric inversion) then $\kmax$ takes its highest possible value (i.e. model top).
The minimum $k$ value $\kmin$ equals 1 if there is no inversion below \kmax, and otherwise is the largest $k< \kmax$ such that $T[k]>T[k-1]$. We then interpolate the column's SW and LW radiative fluxes over this $T$ range onto a uniform $T$ grid running from 150 to 350 K in increments of 2 K, and assign these interpolated profiles, weighted by column area,  to the appropriate \Ts\ bin using $T[1]$ (where \Ts-binning is done with the same uniform grid as for vertical levels $T$). We repeat this for each GCM column over the last 30 years of each simulation, keeping track of the accumulated column area for each bin and $T$ level. This allows us to produce an area-weighted average flux profile in each bin, where in a given bin the total area represented at each $T$ level drops off at lower and higher $T$  (due to small variations in $T[\kmin]$ and $T[\kmax]$ within the bin). These average flux profiles (one per bin) may then be differentiated with respect to $T$, yielding the $-\ppt \Fnet$ profiles shown in Fig. \ref{fnet_ipsl} and \ref{fnet_all}. To reduce noise in these figures, the profiles are cut off once the total area at a given $T$ is less than half of the maximum value in the vertical (where this maximum value is taken throughout most of that bin's tropospheric $T$ range, as expected). 

\begin{figure}[t]
	\begin{center}
			\includegraphics[scale=0.7]{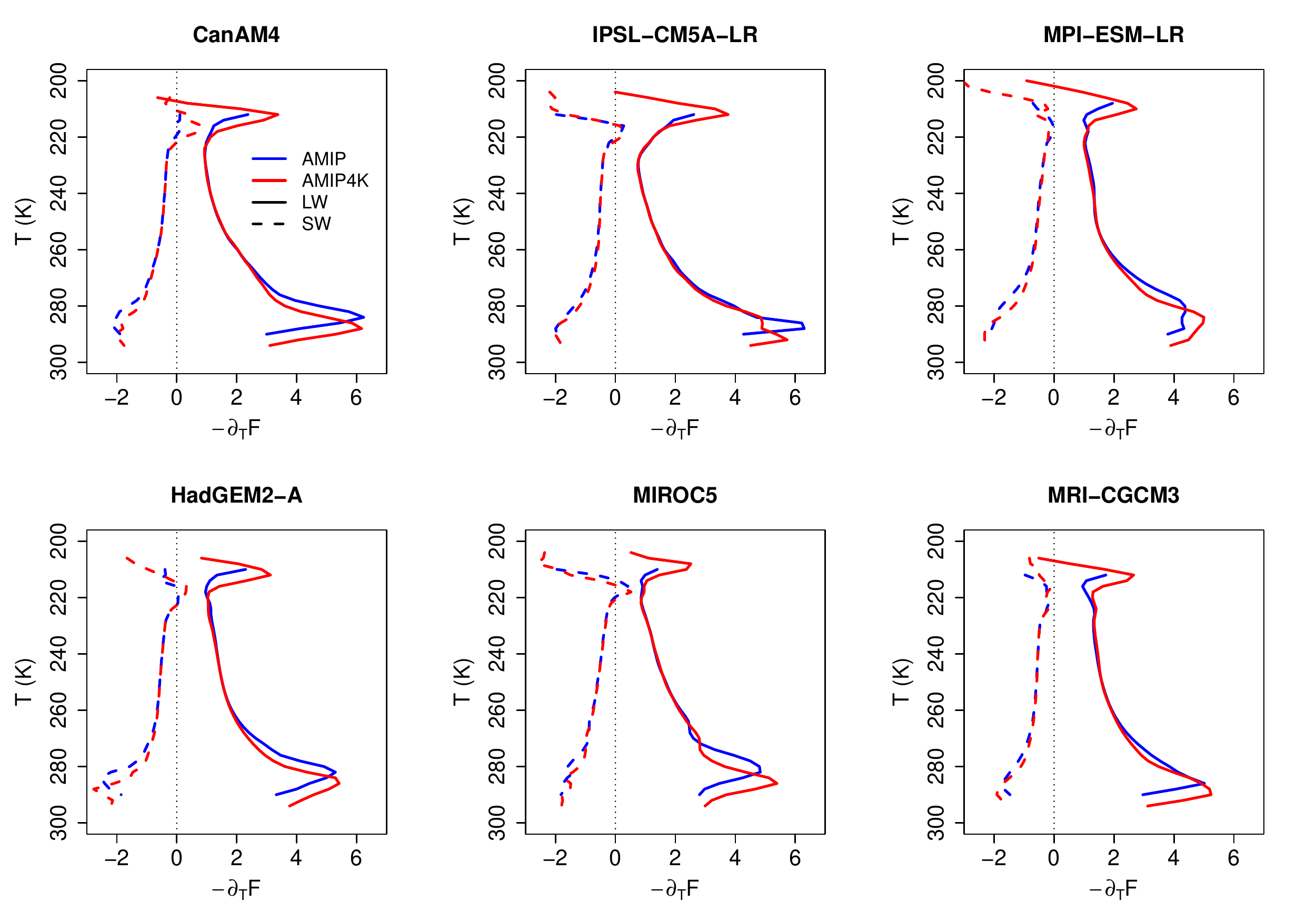}
		\caption{ Profiles of $-\ppt \FSW$ and $-\ppt \FLW$ for the \Ts=290 K (AMIP) and 294 K (AMIP4K) bins for all six CFMIP models. These profiles show that \Ts-invariance in the mid and upper troposphere across models holds for both the LW and SW separately. 
		\label{fswlw_all}
		}
	\end{center}
\end{figure}

The decomposition of these net flux divergence profiles into their LW and SW components is given in Fig. \ref{fswlw_all}. These panels show that mid and upper tropospheric \Ts-invariance holds for the LW and SW separately in the GCMs, just as for the CRM.

\bibliographystyle{apa}

\begin{thebibliography}{}

\bibitem[\protect\astroncite{Allen and Ingram}{2002}]{allen2002}
Allen, M.~R. and Ingram, W.~J. (2002).
\newblock {Constraints on future changes in climate and the hydrologic cycle.}
\newblock {\em Nature}, 419(6903):224--232.

\bibitem[\protect\astroncite{Goldblatt et~al.}{2013}]{goldblatt2013}
Goldblatt, C., Robinson, T.~D., Zahnle, K.~J., and Crisp, D. (2013).
\newblock {Low simulated radiation limit for runaway greenhouse climates}.
\newblock {\em Nature Geoscience}, 6(8):661--667.

\bibitem[\protect\astroncite{Held and Soden}{2006}]{held2006}
Held, I.~M. and Soden, B.~J. (2006).
\newblock {Robust responses of the hydrological cycle to global warming}.
\newblock {\em Journal of Climate}, 19:5686--5699.

\bibitem[\protect\astroncite{Igel et~al.}{2014}]{igel2014}
Igel, A.~L., Igel, M.~R., and van~den Heever, S.~C. (2014).
\newblock {Make it a double: Sobering results from simulations using
  single-moment microphysics schemes.}
\newblock {\em Journal of Atmospheric Sciences}, In Review:910--925.

\bibitem[\protect\astroncite{Ingram}{2010}]{ingram2010}
Ingram, W.~J. (2010).
\newblock {A very simple model for the water vapour feedback on climate
  change}.
\newblock {\em Quarterly Journal of the Royal Meteorological Society},
  136(646):30--40.

\bibitem[\protect\astroncite{Krueger et~al.}{1995}]{krueger1995}
Krueger, S.~K., Fu, Q., Liou, K.~N., and Chin, H.-N.~S. (1995).
\newblock {Improvements of an Ice-Phase Microphysics Parameterization for Use
  in Numerical Simulations of Tropical Convection}.
\newblock {\em Journal of Applied Meteorology}, 34:281--287.

\bibitem[\protect\astroncite{Lambert and Webb}{2008}]{lambert2008}
Lambert, F.~H. and Webb, M.~J. (2008).
\newblock {Dependency of global mean precipitation on surface temperature}.
\newblock {\em Geophysical Research Letters}, 35(16):1--5.

\bibitem[\protect\astroncite{{Le Hir} et~al.}{2009}]{lehir2009}
{Le Hir}, G., Donnadieu, Y., Godd{\'{e}}ris, Y., Pierrehumbert, R.~T.,
  Halverson, G.~P., Macouin, M., N{\'{e}}d{\'{e}}lec, A., and Ramstein, G.
  (2009).
\newblock {The snowball Earth aftermath: Exploring the limits of continental
  weathering processes}.
\newblock {\em Earth and Planetary Science Letters}, 277(3-4):453--463.

\bibitem[\protect\astroncite{Lin et~al.}{1983}]{lin1983}
Lin, Y.-L., Farley, R.~D., and Orville, H.~D. (1983).
\newblock {Bulk Parameterization of the Snow Field in a Cloud Model}.

\bibitem[\protect\astroncite{Lord et~al.}{1984}]{lord1984}
Lord, S.~J., Willoughby, H.~E., and Piotrowicz, J.~M. (1984).
\newblock {Role of a Parameterized Ice-Phase Microphysics in an Axisymmetric,
  Nonhydrostatic Tropical Cyclone Model}.
\newblock {\em Journal of the Atmospheric Sciences}, 41(19):2836--2848.

\bibitem[\protect\astroncite{Margolin et~al.}{2006}]{margolin2006}
Margolin, L.~G., Rider, W.~J., and Grinstein, F.~F. (2006).
\newblock {Modeling turbulent flow with implicit LES}.
\newblock {\em Journal of Turbulence}, 7:N15.

\bibitem[\protect\astroncite{Mlawer et~al.}{1997}]{mlawer1997}
Mlawer, E.~J., Taubman, S.~J., Brown, P.~D., Iacono, M.~J., and Clough, S.~A.
  (1997).
\newblock {Radiative transfer for inhomogeneous atmospheres: RRTM, a validated
  correlated-k model for the longwave}.
\newblock {\em Journal of Geophysical Research}, 102(D14):16663.

\bibitem[\protect\astroncite{Muller et~al.}{2011}]{muller2011b}
Muller, C.~J., O'Gorman, P.~A., and Back, L.~E. (2011).
\newblock {Intensification of precipitation extremes with warming in a
  cloud-resolving model}.
\newblock {\em Journal of Climate}, 24(11):2784--2800.

\bibitem[\protect\astroncite{O'Gorman et~al.}{2012}]{ogorman2012}
O'Gorman, P.~a., Allan, R.~P., Byrne, M.~P., and Previdi, M. (2012).
\newblock {Energetic Constraints on Precipitation Under Climate Change}.
\newblock {\em Surveys in Geophysics}, 33:585--608.

\bibitem[\protect\astroncite{Pendergrass and Hartmann}{2014}]{pendergrass2014}
Pendergrass, A.~G. and Hartmann, D.~L. (2014).
\newblock {The Atmospheric Energy Constraint on Global-Mean Precipitation
  Change}.
\newblock {\em Journal of Climate}, 27(2):757--768.

\bibitem[\protect\astroncite{Pierrehumbert}{1999}]{pierrehumbert1999}
Pierrehumbert, R.~T. (1999).
\newblock {Subtropical water vapor as a mediator of rapid global climate
  change}.
\newblock In Clark, P.~U., Webb, R.~S., and Keigwin, L.~D., editors, {\em
  Mechanisms of Global Climate Change at Millennial}. American Geophysical
  Union, Washington, D. C.

\bibitem[\protect\astroncite{Rodgers and Walshaw}{1966}]{rodgers1966}
Rodgers, C.~D. and Walshaw, C.~D. (1966).
\newblock {The computation of infra-red cooling rate in planetary atmospheres}.
\newblock {\em Quarterly Journal of the Royal Meteorological Society},
  92:67--92.

\bibitem[\protect\astroncite{Romps}{2008}]{romps2008}
Romps, D.~M. (2008).
\newblock {The Dry-Entropy Budget of a Moist Atmosphere}.
\newblock {\em Journal of the Atmospheric Sciences}, 65(12):3779--3799.

\bibitem[\protect\astroncite{Romps}{2011}]{romps2011}
Romps, D.~M. (2011).
\newblock {Response of Tropical Precipitation to Global Warming}.
\newblock {\em Journal of the Atmospheric Sciences}, 68(1):123--138.

\bibitem[\protect\astroncite{Romps}{2014}]{romps2014}
Romps, D.~M. (2014).
\newblock {An Analytical Model for Tropical Relative Humidity}.
\newblock {\em Journal of Climate}, 27(19):7432--7449.

\bibitem[\protect\astroncite{Simpson}{1928}]{simpson1928}
Simpson, G. (1928).
\newblock {Some Studies in Terrestriall Radiation}.
\newblock {\em Memoirs of the Royal Meteorological Society}, 2(16):69--95.

\bibitem[\protect\astroncite{Simpson et~al.}{1988}]{simpson1988}
Simpson, J., Adler, R.~F., North, G.~R., Simpson, J., Adler, R.~F., and North,
  G.~R. (1988).
\newblock {A Proposed Tropical Rainfall Measuring Mission (TRMM) Satellite}.
\newblock {\em
  http://dx.doi.org/10.1175/1520-0477(1988)069{\textless}0278:APTRMM{\textgreater}2.0.CO;2}.

\bibitem[\protect\astroncite{Stephens and Ellis}{2008}]{stephens2008a}
Stephens, G.~L. and Ellis, T.~D. (2008).
\newblock {Controls of Global-Mean Precipitation Increases in Global Warming}.
\newblock {\em Journal of Climate}, pages 6141--6155.

\bibitem[\protect\astroncite{Thomas and Stamnes}{2002}]{thomas2002}
Thomas, G.~E. and Stamnes, K. (2002).
\newblock {\em {Radiative Transfer in the Atmosphere and Ocean}}.
\newblock Cambridge University Press.

\bibitem[\protect\astroncite{Tsushima et~al.}{2015}]{tsushima2015}
Tsushima, Y., Iga, S.~I., Tomita, H., Satoh, M., Noda, A.~T., and Webb, M.~J.
  (2015).
\newblock {High cloud increase in a perturbed SST experiment with a global
  nonhydrostatic model including explicit convective processes}.
\newblock {\em Journal of Advances in Modeling Earth Systems}, 6(3):571--585.

\bibitem[\protect\astroncite{Wing et~al.}{2017}]{wing2017b}
Wing, A.~A., Satoh, M., Reed, K.~A., Stevens, B., and Bony, S. (2017).
\newblock {Radiative-Convective Equilibrium Model Intercomparison Project}.
\newblock {\em Geoscientific Model Development Discussions}, pages 1--24.

\end{thebibliography}

\end{document}